\documentclass[twocolumn,showpacs,showkeys]{revtex4}
\usepackage{graphicx}
\usepackage{bm}
\usepackage{color}
\usepackage{amsmath}
\usepackage{natbib}

\begin{document}

\title{Simultaneous dipole and quadrupole moment contribution in the Bogoliubov spectrum: Application of the non-integral Gross-Pitaevskii equation}

\author{Pavel A. Andreev}
\email{andreevpa@physics.msu.ru}
\affiliation{Faculty of physics, Lomonosov Moscow State University, Moscow, Russian Federation.}

 \date{\today}

\begin{abstract}

We present the Gross-Pitaevskii equation for Bose-Einstein condensates (BECs) possessing the electric dipole and the electric quadrupole moments in a non-integral form. These equations are coupled with the Maxwell equations. The model under consideration includes the dipole-dipole, the dipole-quadrupole, and the quadrupole-quadrupole interactions in terms of the electric field created by the dipoles and quadrupoles. We apply this model to obtain the Bogoliubov spectrum for three dimensional BECs with a repulsive short-range interaction. We obtain an extra term in the Bogoliubov spectrum in compare with the dipolar BECs. We show that the quadrupole-quadrupole interaction gives a positive contribution in the Bogoliubov spectrum. Hence this spectrum is stable.
\end{abstract}

%Use showkeys class option if keyword

\maketitle

%%%%%%%%%%TEXT

\section{Introduction}

On the steady interest to dipolar quantum gases \cite{Ni PCCP 09}-\cite{Adhikari JP B 14}, there has arisen interest to quantum gases with the quadrupole moments \cite{Yongyao Li PRA 13}-\cite{Picovski PRA 14}. Earlier papers on this subject are focused on quantum phases of quadrupolar Fermi gases located in different traps \cite{Wen-Min Huang PRA 14}, \cite{Bhongale PRL 13}. A non-linear Schrodinger equation was applied to consider solitons in quantum gases with the long-range
quadrupole-quadrupole interaction \cite{Yongyao Li PRA 13}. An explicit form of the potential energy of the
quadrupole-quadrupole interaction was used in Ref. \cite{Yongyao Li PRA 13}. It results in an integral form of the Gross-Pitaevskii equation (see formula (10) in Ref. \cite{Yongyao Li PRA 13}).  It was demonstrated that some aspects of cold gases with
anisotropic interparticle interactions can be studied in a general way, with very
little assumptions on the form for the interparticle interactions \cite{Picovski PRA 14}.

Efforts to explore of quadrupolar quantum gases are encouraged %inspired
by experimental data on the quadrupole moment of atoms and molecules \cite{Buckingham JACS 68}-\cite{Byrd JCP 11}.

Structure of external fields for creation of the electric and magnetic quadrupoles built as tightly bound pairs of dipoles with orientations opposite to each other is described in Ref. \cite{Yongyao Li PRA 13}.

In our paper we are focused on the electric quadrupolar particles being in the Bose-Einstein condensate (BEC) state. We assume that objects have both the quadrupole electric moment (QEM) and the dipole electric moment (DEM).

We assume that all dipoles are aligned. We also assume that all quadrupoles have same magnitude and tensor structure. Hence we have system of particles moving without any particle deformations or oscillations of the particle dipole directions. Consequently the evolution of the dipole and quadrupole electric moments reduces
to the evolution of the particle concentration. Nevertheless, the concentration evolution are affected by the dipole-dipole, the quadrupole-dipole, and the quadrupole-quadrupole interactions. These interactions are presented in the Euler equation via corresponding force fields.

Applying all assumptions described above we obtain the set of quantum hydrodynamic equations consisting of the continuity and Euler equations. We consider the dipole-dipole, the dipole-quadrupole, the quadrupole-quadrupole interactions as long-range interactions. We could apply the self-consistent field approximation, but it cannot be done for particles in the BEC state (see discussion before formula (\ref{QDbec n2 expansion}) of this paper, see also Ref. \cite{Andreev Ar_TW_14}). It appears that the long-range interactions of the particles in the BEC state have the exchange part only, and the self-consistent field equals to zero \cite{Andreev Ar_TW_14}. Fortunately the exchange part arises in the form coinciding with the formal application of the self-consistent field approximations. The last one was considered in Refs. \cite{Andreev 2013 non-int GP}, \cite{Andreev 2013 Dip+Spin}, \cite{Andreev EPJ D Pol}. After all we obtain the non-integral quantum hydrodynamic equations. These equations appear together with the equations of field, which are the pair of the quasi-electrostatic Maxwell equations. The densities of the DEM and QEM came in the Maxwell equations as the sources of the potential electric field.

Under the assumption of potential velocity field we derive corresponding non-integral non-linear Schrodinger equation, which is the generalization of the Gross-Pitaevskii equation for particles possessing the DEM and the QEM.

The non-integral form of the Gross-Pitaevskii equation for electrically dipolar BECs was obtained in Ref. \cite{Andreev 2013 non-int GP}. There were also presented corresponding quantum hydrodynamic equations. The electric field created by the dipoles was explicitly considered there. The electric field evolution in dipolar BECs was also considered in Ref. \cite{Wilson NJP 12}. The non-integral description of magnetically dipolar BECs was presented in Ref. \cite{Andreev 2013 Dip+Spin}. Differences in behavior of the align electric dipoles and the align magnetic dipoles was demonstrated in Ref. \cite{Andreev 2013 Dip+Spin}. A generalization of described results for finite size particles was performed in Ref. \cite{Andreev 2014 finite size BEC}. The model of dipolar BECs with the dipole direction evolution was developed in Ref. \cite{Andreev EPJ D Pol}. These researches have created a background for this paper.

The non-integral equation for the collective particle evolution under influence of a long-range interaction appears together with equations of field. The Maxwell equations are equations of field for the electromagnetic field. Hence we present a description of dipolar BECs, which is in agreement with the Maxwell electrodynamics \cite{Landau 2}.

This brief paper is organized as follows. In Sec. II we present equation of quantum hydrodynamics and generalized non-integral Gross-Pitaevskii equation for BECs with the DEM and the QEM. In Sec. III we calculate the Bogoliubov spectrum for the small amplitude collective excitations. In Sec. IV a summary of obtained results is presented.

\section{Model}

The method of many-particle quantum hydrodynamics \cite{MaksimovTMP 1999} allows to derive the Gross-Pitaevskii equation for BECs of neutral atoms \cite{Andreev PRA08}. The Gross-Pitaevskii equation appears in the first order by the interaction radius. A generalization of this model appears at more detail consideration of the short-range interaction up to the third order by the interaction radius Ref. \cite{Andreev PRA08}. Corresponding generalization of the Bogoloubov spectrum can be found in Ref. \cite{Andreev PRA08}.

The method of many-particle quantum hydrodynamics proves to be useful at consideration of the three-particle interaction in BECs and ultra-cold Bose atoms at non-zero temperatures \cite{Andreev IJMP B 13}.

Different long-range interactions have been also considered in terms of the many-particle quantum hydrodynamics \cite{Andreev 2013 Dip+Spin}, \cite{Andreev EPJ D Pol}, \cite{MaksimovTMP 1999}, \cite{Andreev PRB 11}, including the electric dipole \cite{Andreev 2013 non-int GP}, \cite{Andreev EPJ D Pol}, \cite{Andreev PRB 11}, \cite{Andreev JP B 14} and magnetic dipole (the spin-spin) \cite{Andreev 2013 Dip+Spin} interactions.

We consider now BECs with the DEM and the QEM.

We obtain that motion of the medium obeys the following quantum hydrodynamic equations
\begin{equation}\label{QDbec continuity equation} \partial_{t}n+\nabla (n\textbf{v})=0,
\end{equation}
and
$$mn(\partial_{t}+\textbf{v}\nabla)\textbf{v}-\frac{\hbar^{2}}{4m}n\nabla\Biggl(\frac{\triangle n}{n}-\frac{(\nabla n)^{2}}{2n^{2}}\Biggr)$$
\begin{equation}\label{QDbec Euler integral} =-gn\nabla n+d l^{\beta}n\nabla E^{\beta}+\frac{1}{6}Qq^{\beta\gamma}n\nabla \partial^{\gamma}E^{\beta},
\end{equation}
where $n$ is the particle concentration, $\textbf{v}$ is the velocity field, $\partial_{t}$ is the time derivative, $\nabla$ and $\partial^{\alpha}$ are the gradient operator consisting of partial spatial derivatives, $\triangle$ is the Laplace operator, $m$ is the mass of particles, $\hbar$ is the reduced Planck constant, $g$ is the interaction constant for the short-range interaction, $d$ ($Q$) is the magnitude of dipole (quadrupole) electric moment, $\textbf{l}$ is the vector showing equilibrium direction of the dipoles, $q^{\alpha\beta}$ is the second rank tensor showing structure of the quadrupole moment of particles.

Equation (\ref{QDbec continuity equation}) is the continuity equation showing conservation of the particle number. Equation (\ref{QDbec Euler integral}) is the Euler equation, which is the momentum balance equation. The group of terms, on the left-hand side of the Euler equation (\ref{QDbec Euler integral}), proportional to the square of the Planck constant $\hbar^{2}$, is the quantum Bohm potential \cite{Bahm PR 53}. The right-hand side of the Euler equation consists of three terms presenting different interactions. The first term describes the short-range interaction in the Gross-Pitaevskii approximation, or, in other words, in the first order by the interaction radius \cite{Andreev PRA08}, \cite{L.P.Pitaevskii RMP 99}. The second (third) term presents action of the electric field created by the DEM and QEM (internal electric field) on the DEM (QEM) density. Hence the second term contains the dipole-dipole and part of dipole-quadrupole interaction. The third term contains another part of the dipole-quadrupole interaction and the quadrupole-quadrupole interaction.

Comprehensive analysis of approximations of quantum hydrodynamic description of dipolar BECs was presented in Ref. \cite{Andreev Ar_TW_14}. It shows that the dipole-dipole, quadrupole-dipole, and quadrupole-quadrupole interactions between particles in the BEC state, presented in the Euler equation (\ref{QDbec Euler integral}) correspond to the exchange part of these interactions, instead of the self-consistent field part as it had been expected earlier \cite{Andreev 2013 Dip+Spin}, \cite{Andreev EPJ D Pol}. However, the explicit form of exchange correlations for bosons in the BEC state is in formal coincidence with the self-consistent field terms. Hence it allows to introduce the electric field created by dipoles and quadrupoles as we demonstrate it below.

For instance the force field of dipole-dipole interaction in the general Euler equation appears as $\textbf{F}(\textbf{r},t)= -\int d\textbf{r}'(\nabla U_{dd}) n_{2}(\textbf{r},\textbf{r}',t)$, with the potential energy of dipole-dipole interaction $U_{dd}=-d^{\beta}d^{\gamma}G^{\beta\gamma}(|\textbf{r}-\textbf{r}'|)$. The self-consistent field term related to the part of the two particle concentration describing interaction of pairs of particles being in different quantum states and reveals in the simple representation of the two-particle concentration as the product of one-particle concentrations $n_{2}(\textbf{r},\textbf{r}',t)\rightarrow n(\textbf{r},t) n(\textbf{r}',t)$ (see the first term on the right-hand side of formula 4 in Ref. \cite{Andreev PRA08}, see formula 25 and discussions after formula 23 in Ref. \cite{Andreev PRA08}). However all ultracold bosons are located in the single quantum state with the lowest energy. The self-consistent field equals to zero for these particles. Thus we need to consider full formula for the two-particle concentration of weakly interacting bosons, which contains the exchange interaction contribution. This formula was derived in Ref. \cite{Andreev PRA08} (see formula (25)). We also present it here
$$n_2(\textbf{r},\textbf{r}',t)=n(\textbf{r},t)n(\textbf{r}',t)$$
\begin{equation}\label{QDbec n2 expansion}
+|\rho(\textbf{r},\textbf{r}',t)|^{2}+\sum_{g}n_{g}(n_{g}-1)|\varphi_{g}(\textbf{r},t)|^{2}|\varphi_{g}(\textbf{r}',t)|^{2},\end{equation}
where
\begin{equation}\label{QDbec rho varphi}\rho(\textbf{r},\textbf{r}',t)=\sum_{g}n_{g}\varphi_{g}^{*}(\textbf{r},t)\varphi_{g}(\textbf{r}',t),\end{equation}
and $\varphi_{g}(\textbf{r},t)$ are the arbitrary single-particle wave functions. We can also present the particle concentration in terms of single-particle wave functions $n(\textbf{r},t)=\sum_{g}n_{g}\varphi_{g}^{*}(\textbf{r},t)\varphi_{g}(\textbf{r},t)$. The second and third terms in formula (\ref{QDbec n2 expansion}) are related to the exchange interaction. The second term describes exchange interaction of bosons located in different quantum states, and the third term presents the exchange interaction of bosons being in the same quantum state with quantum numbers $g$, with summation over all quantum states. Only one term in expansion of the two-particle concentration (\ref{QDbec n2 expansion}) survives for particles in the BEC state. This is the term of sum corresponding to the quantum state with lowest energy $g_{0}$. Consequently the two-particle concentration can be written as $n_2(\textbf{r},\textbf{r}',t)=n_{g}|\varphi_{g}(\textbf{r},t)|^{2}\cdot n_{g}|\varphi_{g}(\textbf{r}',t)|^{2}$$=n(\textbf{r},t)n(\textbf{r}',t)$, where we have assumed $n_{g_{0}}\gg1$. Behavior of the exchange correlations in ultracold fermions is rather different, analysis of the Coulomb exchange interaction in quantum plasmas of degenerate electrons was presented in Ref. \cite{Andreev 1403 exchange}.

The internal electric field consists of two parts. One of them is created by electric dipoles. Its explicit form is $E^{\alpha}_{dip}(\textbf{r},t)=\int d\textbf{r}' G^{\alpha\beta}(\textbf{r},\textbf{r}')P^{\beta}(\textbf{r}',t)$, with the Green function of electric dipole interaction $G^{\alpha\beta}(\textbf{r},\textbf{r}')=\partial^{\alpha}\partial^{\beta}\frac{1}{\mid\textbf{r}-\textbf{r}'\mid}$. Structure of the polarization for the align dipoles is $\textbf{P}(\textbf{r},t)=d\textbf{l} n(\textbf{r},t)$, where $\textbf{l}$ is the fixed direction of the dipoles. The second part of the internal electric field is created by the electric quadrupoles $E^{\alpha}_{quad}(\textbf{r},t)=-\frac{1}{6}\int d\textbf{r}'G^{\alpha\beta\gamma}(\textbf{r},\textbf{r}')Q^{\beta\gamma}(\textbf{r}',t)$, where $G^{\alpha\beta\gamma}(\textbf{r},\textbf{r}')=\partial^{\alpha}\partial^{\beta}\partial^{\gamma}\frac{1}{\mid\textbf{r}-\textbf{r}'\mid}$ is the Green function giving the electric field created by the quadrupoles, $Q^{\alpha\beta}(\textbf{r},t)=q^{\alpha\beta} Q n(\textbf{r},t)$ is the density of quadrupoles moving without any deformation of particles or oscillation of the direction of particle symmetry axes, $q^{\alpha\beta}$ is the unit tensor showing the tensor structure of QEM. Sum of these fields satisfies the Maxwell equations
\begin{equation}\label{QDbec div E} \nabla\cdot \textbf{E}(\textbf{r},t)=-4\pi \biggl(dl^{\alpha} \partial^{\alpha}-\frac{1}{6}Qq^{\alpha\beta}\partial^{\alpha}\partial^{\beta}\biggr)n(\textbf{r},t),
\end{equation}
and
\begin{equation}\label{QDbec curl E} \nabla\times \textbf{E}(\textbf{r},t)=0.
\end{equation}
Equation (\ref{QDbec div E}) is the Poisson equation. Densities of electric dipoles and electric quadrupoles come in the right-hand side of the Poisson equation. To give the Poisson equation the traditional form we can introduce an effective polarization $\textbf{P}_{eff}$ containing contribution of electric dipole moments and electric quadrupole moments $P_{eff}^{\alpha}=dl^{\alpha}n-\frac{1}{6}Qq^{\alpha\beta}\partial^{\beta}n$. Then equation (\ref{QDbec div E}) can be written as $\nabla\cdot \textbf{E}(\textbf{r},t)=-4\pi\nabla\cdot \textbf{P}_{eff}$. The self-consistent field approximation for system of particles with a long-range interaction was suggested by Vlasov in 1938 \cite{Vlasov JETP 38}. However we have not applied this approximation in equation (\ref{QDbec Euler integral}), but we consider the exchange dipole-dipole, quadrupole-dipole, and quadrupole-quadrupole interactions appearing in the form coinciding with the self-consistent field.

Since we consider the quasi-static evolution of dipoles the Euler equation (\ref{QDbec Euler integral}) is coupled with the Poisson equation (\ref{QDbec div E}), instead of the full set of Maxwell equations. Hence obtained equations (\ref{QDbec continuity equation})-(\ref{QDbec curl E}) correspond the Vlasov-Poisson approximation, which is an analog of the well-known model in the plasma physics.

The system of hydrodynamic equations (\ref{QDbec continuity equation}), (\ref{QDbec Euler integral}) can be replaced by the generalized non-integral Gross-Pitaevskii equation for the macroscopic wave function $\Phi(\textbf{r},t)$
$$\imath\hbar\partial_{t}\Phi(\textbf{r},t)=\Biggl(-\frac{\hbar^{2}}{2m}\triangle+g\mid\Phi(\textbf{r},t)\mid^{2}$$
\begin{equation}\label{QDbec nlse polariz non Int ED} -d\textbf{l}\textbf{E}(\textbf{r},t)-\frac{1}{6}Q^{\alpha\beta}(\partial^{\alpha}E^{\beta})\Biggr)\Phi(\textbf{r},t),\end{equation}
with
\begin{equation}\label{QDbec WF def} \Phi(\textbf{r},t)=\sqrt{n(\textbf{r},t)}\exp(\imath m\phi(\textbf{r},t)/\hbar),\end{equation}
where $\phi(\textbf{r},t)$ is the potential of velocity field $\textbf{v}(\textbf{r},t)=\nabla\phi(\textbf{r},t)$.

Equation (\ref{QDbec nlse polariz non Int ED}) is coupled with the Maxwell equations (\ref{QDbec div E}) and (\ref{QDbec curl E}) via the electric field $\textbf{E}$. The particle concentration $n$ is related to the macroscopic wave function $\Phi(\textbf{r},t)$ in the usual way $n=|\Phi|^{2}$.

\section{Bogoliubov spectrum}

Considering the small amplitude perturbations of the equilibrium state $\delta n=N\exp(-i\omega t+i\textbf{k}\textbf{r})$, $\delta \textbf{v}=\textbf{U}\exp(-i\omega t+i\textbf{k}\textbf{r})$,  and $\delta \textbf{E}=\mbox{\boldmath $\varepsilon$}\exp(-i\omega t+i\textbf{k}\textbf{r})$, with $\textbf{k}=\{k_{x}, k_{y}, k_{z}\}$, we can obtain spectrum $\omega(\textbf{k})$. $N$, $\textbf{U}$ and $\mbox{\boldmath $\varepsilon$}$ are constant amplitudes of oscillations.

A quadrupole at zero dipole moment is a pair of two dipoles, which have equal module and opposite directions. An object with quadrupole moment and non-zero dipole moment is similar to previous case, but dipoles of the pair have different length. Let us introduce notation $d_{in}$ for shorter dipole of the pair (the inner dipole), and notation $d_{b}$ for larger dipole of the pair, and a notation for difference of them $d_{out}=d_{b}-d_{in}$ (the outer dipole). So, the outer dipole $d_{out}$ forms the dipole-dipole interaction, then the inner dipole forms the quadrupole moment of particles. Considering electric quadrupole moment as a tight pair of antiparallel electric dipoles with magnitude $d_{in}$ being separated by distance $\epsilon$ we have $Q=3d_{in} \epsilon$ \cite{Yongyao Li PRA 13}. We assume that dipoles parallel to the z axis, and the interdipole distance is parallel to the x axis.

We have that the equilibrium polarization is directed parallel $z$ axes. Tensor of equilibrium quadrupole moment $q^{\alpha\beta}$ has the following structure: $q^{xz}=q^{zx}=1$, $q^{xx}=q^{yy}=q^{zz}=q^{yz}=q^{xy}=0$.

Linearised set of QHD equations (\ref{QDbec continuity equation}), (\ref{QDbec Euler integral}), after substituting of monochromatic perturbations gives the following relations between $N$, $\textbf{U}$ and $\varepsilon_{x}$, $\varepsilon_{z}$: $N=n_{0}\frac{\textbf{k}\textbf{U}}{\omega}$, and
\begin{equation}\label{QDbec kU=} \textbf{k}\textbf{U}=\frac{-\omega k^{2}[d\varepsilon_{z} +\frac{1}{6}Q\imath(k_{x}\varepsilon_{z}+k_{z}\varepsilon_{x})]}{m\omega^{2}-\frac{\hbar^{2}k^{4}}{4m}-gn_{0}k^{2}}.\end{equation}

Imaginary part of the numerator in formula (\ref{QDbec kU=}) demonstrates that some kind of instability can arise from the electric dipole-quadrupole interaction. However the Poisson equation (\ref{QDbec div E}) contains contribution of "dipole-quadrupole" and "quadrupole-dipole" interactions canceling each other. Hence final spectrum has no imaginary part.

The equation of field (\ref{QDbec curl E}) leads to
\begin{equation}\label{QDbec} \begin{array}{cc}
                                  \varepsilon_{x}=\frac{k_{x}}{k_{z}}\varepsilon_{z}, & \varepsilon_{y}=\frac{k_{y}}{kz}\varepsilon_{z}.  \\
                                \end{array}
\end{equation}

All these intermediate results can be substituted in linearised Poisson equation (\ref{QDbec div E}). Before substituting of $N$, $\varepsilon_{x}$, $\varepsilon_{y}$ in the Poisson equation its linearised form appears as
\begin{equation}\label{QDbec}  k_{x}\varepsilon_{x} +k_{y}\varepsilon_{y} +k_{z}\varepsilon_{z} =-4\pi\biggl(dk_{z}\delta n -\frac{1}{3}Q\imath k_{x}k_{z}\delta n\biggr), \end{equation}
where we have included $q^{\alpha\beta}k^{\alpha}k^{\beta}$ $=q^{xz}k_{x}k_{z}+q^{zx}k_{z}k_{x}$ $=2k_{x}k_{z}$.

After the substituting we obtain
\begin{equation}\label{QDbec}  k^{2}\varepsilon_{z}=\frac{4\pi k_{z}^{2}n_{0}[d^{2}+\frac{1}{9}Q^{2}k_{x}^{2}]k^{2}\varepsilon_{z}}{m\omega^{2}-\frac{\hbar^{2}k^{4}}{4m}-gn_{0}k^{2}},\end{equation}
where we have included that two terms proportional to the product of the electric dipole and electric quadrupole moments $dQ$ cancel each other.

\begin{figure}
\includegraphics[width=8cm,angle=0]{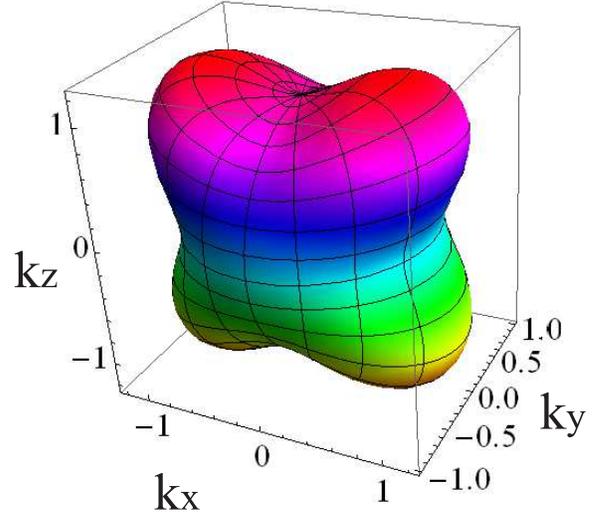}
\caption{\label{QDbec fig 1} (Color online) Parametric form of the angular dependence of the group of terms proportional to $k^{4}$, containing the quadrupolar term, of the Bogoliubov spectrum (\ref{QDbec disp for discuss general with angles}) is presented on the figure. This group of terms dominates in the spectrum in the short-wavelength limit. We apply the following parameters of particles: mass $m=2.85$ $10^{-22}$g, particle concentration $n_{0}=10^{14}$ cm$^{-3}$, and quadrupole moment $Q=50$ Debye {\AA}.}
\end{figure}

Our calculation gives the following spectrum of collective excitations
$$\omega^{2}=\frac{\hbar^{2}k^{4}}{4m^{2}}+\frac{gn_{0}k^{2}}{m}+\frac{4\pi n_{0}d^{2}k^{2}}{m}\cos^{2}\theta$$
\begin{equation}\label{QDbec disp for discuss general with angles} +\frac{1}{9}\frac{4\pi n_{0}Q^{2}}{m}k^{4}\cos^{2}\theta\sin^{2}\theta\cos^{2}\varphi.\end{equation}
where $k_{x}=k\sin\theta\cos\varphi$, $k_{z}=k\cos\theta$, and $k=\sqrt{k_{x}^{2}+k_{y}^{2}+k_{z}^{2}}$, $k_{y}$ appears inside $k$ only. This is the spectrum of longitudinal waves in dipolar-quadrupolar BECs.

We have four positive terms in spectrum (\ref{QDbec disp for discuss general with angles}). The first of these terms is the contribution of the quantum Bohm potential. The second term is the short-range interaction contribution considered in the first order by the interaction radius \cite{Andreev PRA08}, \cite{L.P.Pitaevskii RMP 99}. The third term is the electric dipole moment  contribution \cite{Andreev 2013 non-int GP}, \cite{Andreev 2013 Dip+Spin}. The last term is the contribution of the quadrupole electric moment in the Bogoliubov spectrum. This term is one of main results of our paper.

The electric dipole interaction gives a shift of the short-range interaction constant $g\rightarrow g+4\pi d^{2}\cos^{2}\theta$ giving an anisotropic spectrum. The quadrupole electric moment gives an anisotropic shift of the quantum Bohm potential contribution $\frac{\hbar^{2}}{4m^{2}}\rightarrow\frac{\hbar^{2}}{4m^{2}}+\frac{4\pi n_{0}Q^{2}}{9m}\cos^{2}\theta\sin^{2}\theta\cos^{2}\varphi$.

Studying the dipolar BECs, researchers measure the rate of anisotropy $\eta$ of the spectrum as the ratio between the frequency at propagation of waves parallel to the external electric field $\theta=0$, and the frequency of perturbations propagating perpendicular to the external field $\theta=\pi/2$. Hence we have for dipolar anisotropy $\eta_{D}=\omega_{0}/\omega_{\pi/2}$. As we see from formula (\ref{QDbec disp for discuss general with angles}) and figure (\ref{QDbec fig 1}), the presence of the quadrupolar term in the Bogoliubov spectrum (\ref{QDbec disp for discuss general with angles}) does not change this ratio at any quadrupole moments of particles due to different form of quadrupolar anisotropy. Quadrupolar part $\Omega_{Q}^{2}$ of the spectrum (\ref{QDbec disp for discuss general with angles}) equals to zero at $\theta=0$, and $\theta=\pi/2$. Maximal contribution of the quadrupolar part arises at $\theta=\pi/4$ and $\theta=3\pi/4$, if $\varphi=0$, or $\varphi=\pi$. These complicate conditions for maximum of $\Omega_{Q}^{2}$ arises due to the rather complicate anisotropy of $\Omega_{Q}^{2}(\theta,\varphi)$.

The difference of anisotropy form of the quarupole interaction from the anisotropy of the dipole interaction shows in no modifications of the bright, dark, and grey soliton solutions propagating in electric dipolar-quadrupolar BECs parallel or perpendicular to the equilibrium polarisation considered in Refs. \cite{Andreev 2013 non-int GP} and \cite{Andreev 2013 Dip+Spin}.

Quadrupole moment of particles leads to two characteristic lengths. One of them is given by ratio of quadrupolar part to the short-range interaction part $l_{Q}=\sqrt{\frac{4\pi Q^{2}}{9g}}$. The another one appears from ratio of quadrupolar and dipolar parts $l_{Qd}=\frac{Q}{3d}=\frac{d_{in}}{d_{out}}\epsilon$. $l_{Qd}$ can be larger or smaller than the interdipole distance inside quadrupole $\epsilon$.

Comparison of $l_{Qd}$ (or $l_{Q}$, if the short-range interaction dominates over the dipole-dipole interaction) with the wavelength $\lambda=2\pi/k$ allows to introduce regime where the quadrupole interaction is comparable with the dipole-dipole interaction. Let us make some estimations to compare the quadrupolar and dipolar parts of the spectrum for some common parameters. For quadrupole moment $Q=20$ Debye {\AA} and electric dipole moment $d=0.1$ Debye, we have $l_{Qd}=0.67$ $10^{-6}$ cm. At particle concentration $n_{0}=10^{14}$ cm$^{-3}$ the average interparticle distance $a_{0}=1/\sqrt[3]{n_{0}}=0.4$ $10^{-5}$ cm. Wavelengths $\lambda$ are larger than the average interparticle distance $a_{0}<\lambda$. We can see that at the average electric dipole moment, and rather large electric quadrupole moment, the quadrupolar term is about one percent of the dipolar term, or smaller. Nevertheless we can expect to have rather larger contribution of the quadrupole-quadrupole interaction in denser mediums, where excitations with smaller wavelength can propagate.

Quadrupoles can be built as tightly bound pairs of dipole and anti-dipoles. In this case we do not have dipolar part of the spectrum. Applying the Feshbach resonance we can decrease the interaction constant, so that the quantum Bohm potential and the quadrupolar parts of the spectrum will dominate. In this regime it is essential to compare these two terms.

For particles of mass $m=4.3$ $10^{-23}$g and quadrupole moment $Q=6$ Debye {\AA} (acetylene molecules \cite{Junquera-Hernandez CPL 02}) at the particle concentration $n_{0}\sim10^{14}$ cm$^{-3}$ quadrupolar term is about 0.001 of the quantum Bohm potential. Alkaline diatoms have larger mass and larger electric quadrupole moment ($\sim 50$ Debye {\AA} \cite{Byrd JCP 11}). Hence, at concentration $n_{0}\sim10^{14}$ cm$^{-3}$ the quadrupolar term is about of $0.1\div 1$ of the quantum Bohm potential. The figure (1) shows $\Omega^{2}=\frac{4m^{2}\omega^{2}}{\hbar^{2}k^{4}}=1+2.2\cos^{2}\theta\sin^{2}\theta\cos^{2}\varphi$ for particles of mass $m=2.85$ $10^{-22}$g and quadrupole moment $Q=50$ Debye {\AA}.

In this paper we are interested in the BECs of particles possessing the dipole and quadrupole moments simultaneously. Spectrum (\ref{QDbec disp for discuss general with angles}) shows that the quadrupolar term change the short-wavelength part of spectrum, then main contribution of the dipolar term is in the long-wavelength limit.

\section{Conclusion}

We have developed the method of many-particle quantum hydrodynamics for BECs of particles having the DEM and the QEM. We have obtained the non-integral continuity and Euler equations and corresponding Gross-Pitaevskii equation containing  the electric field created by the DEMs and the QEMs of the medium. This electric field obeys the Maxwell equations. We have derived the Bogoliubov spectrum containing contribution of the dipole-dipole, the quadrupole-dipole and the quadrupole-quadrupole interactions. We have found that the quadrupole-quadrupole interaction gives highly anisotropic positive contribution in the spectrum $\omega^{2}(k)$. The quadrupole-dipole interaction gives no contribution in the Bogoliubov spectrum.

Obtained set of QHD equations and corresponding non-integral Gross-Pitaevskii equation open possibilities for studying of different collective phenomena in quadrupolar BECs and BECs of particles possessing the DEM and the QEM simultaneously.

%%%%%%%%%%%%%%%%%%%%%%%%%%%%%%%%%%%%%%%%%%%%%

\begin{acknowledgements}
The author thanks Professor L. S. Kuz'menkov for fruitful discussions.
\end{acknowledgements}

\end{document}